\documentclass[twocolumn,twocolappendix,tighten,times,astrosymb]{aastex631}

\usepackage{amsmath}
\usepackage{booktabs}
\usepackage{soul}
\usepackage{microtype}
\usepackage[english]{babel}
\usepackage{cancel}
\usepackage{soul}
\hypersetup{linkcolor=blue,citecolor=blue,filecolor=blue,urlcolor=blue}

\newcommand{\be}{\begin{equation}}
\newcommand{\ee}{\end{equation}}

\long\def\exclude#1{}

\shorttitle{Neutrinos from AGN coronae}
\shortauthors{}
\begin{document}

\title{
\large The contribution of turbulent AGN coronae to the diffuse neutrino flux\\
}

\correspondingauthor{damianofg@gmail.com}
\author[0000-0003-4927-9850]{Damiano F. G. Fiorillo}
\affiliation{Deutsches Elektronen-Synchrotron DESY, Platanenallee 6, 15738 Zeuthen, Germany}
\author[0000-0001-8822-8031]{Luca Comisso}
\affiliation{Department of Physics, Columbia University, New York, NY 10027, USA}
\affiliation{Department of Astronomy, Columbia University, New York, NY 10027, USA}
\author[0000-0003-0543-0467]{Enrico Peretti}
\affiliation{INAF - Astrophysical Observatory of Arcetri, Largo E. Fermi 5,
50125 Florence, Italy}
\affiliation{Université Paris Cité, CNRS, Astroparticule et Cosmologie, 10 Rue Alice Domon et Léonie Duquet, F-75013 Paris, France}
\author[0000-0001-6640-0179]{Maria Petropoulou}
\affiliation{Department of Physics, National and Kapodistrian University of Athens, University Campus Zografos, GR 15784, Athens, Greece }
\affiliation{Institute of Accelerating Systems \& Applications, University Campus Zografos, Athens, Greece}
\author[0000-0002-1227-2754]{Lorenzo Sironi}
\affiliation{Department of Astronomy and Columbia Astrophysics 
Laboratory, Columbia University, New York, NY 10027, USA}
\affiliation{Center for Computational Astrophysics, Flatiron Institute, 162 5th Avenue, New York, NY 10010, USA}

\begin{abstract}
Active galactic nuclei (AGN) can accelerate protons to energies of $\sim$ 10--100~TeV, with secondary production of high-energy neutrinos. If the acceleration is driven by magnetized turbulence, the main properties of the resulting proton and neutrino spectra can be deduced based on insights from particle-in-cell simulations of magnetized turbulence.
We have previously shown that these properties are consistent with the TeV~neutrino signal observed from the nearby active galaxy NGC~1068. In this work, we extend this result to a population study. We show that the produced neutrino flux depends mainly on the energetics of the corona -- the relative fraction of X-ray, magnetic, and non-thermal proton energy -- and on the spectral energy distribution of the AGN. We find that coronae with similar properties can explain neutrinos from the candidate AGN for which IceCube has reported an excess, albeit less significant than NGC~1068. Building on this framework, we show how the neutrino signal evolves with the AGN luminosity, and use this AGN sequence to predict the diffuse neutrino flux from the extragalactic population, showing that it can account for the diffuse neutrino signal observed by IceCube in the $\sim$1--100~TeV energy range.
\end{abstract}


\section{Introduction}

The origin of the high-energy astrophysical neutrino flux observed at Earth remains one of the central questions in multi-messenger astrophysics. Since the first detection of these neutrinos by the IceCube collaboration in 2013 \citep{IceCube2013}, spanning an energy range from 10~TeV to 10~PeV, the neutrino spectrum has been characterized with increasing precision. There are now many features that suggest its potential origin not from a single population of astrophysical sources, but rather from multiple populations. These include the first hints that the spectrum is not a single power law, but rather a soft spectrum in the TeV range that hardens at higher energies. The spectrum is characterized by a large neutrino flux at $\sim~10$~TeV, which points towards a class of gamma-ray opaque sources dominating emission in this energy region~\citep{Murase:2013rfa,Murase:2015xka,Capanema:2020rjj,Capanema:2020oet}. 
Such a hypothesis has been formulated based on the possible impact of gamma rays produced in optically thin environments on the diffuse gamma-ray flux observed by Fermi-LAT \citep{Fermi-LAT_Diffuse}. In particular, multi-TeV photons produced in extra-galactic sources are expected to be reprocessed to the sub-TeV range throughout the electromagnetic cascade \citep{Berezinsky_cascade} on the Cosmic Microwave Background (CMB) and the Extragalactic Background Light (EBL) \citep{Franceschini_EBL}, which can be constrained with the Fermi-LAT measurements. Thus, gamma-rays must ultimately be attenuated within the sources of neutrinos in this energy range.

A complementary source of information is the identification of specific candidates for point sources of neutrinos. A milestone in this direction is the IceCube discovery of an excess of neutrinos in the $\sim$1-10
~TeV energy range from the direction of the Seyfert galaxy NGC~1068. This galaxy contains an active galactic nucleus (AGN), namely a compact region with powerful broadband electromagnetic emission associated with accretion onto the central supermassive black hole (SMBH) \citep{Padovani:2024ibi}. The neutrino luminosity inferred by the IceCube observation is in the range $10^{42-43}$~ erg/s. On the other hand, no comparable gamma-ray luminosity has been observed in the TeV range~\citep{MAGIC-UL-NGC1068}, which points towards a radiatively compact region for neutrino production which would be opaque to $\gamma\gamma$ absorption. 
More recently, IceCube has reported from less significant neutrino excesses from the direction of other Seyfert galaxies~\citep{Neronov_Seyfert,IceCube24_Seyfert}, which reinforces the hint towards a sizable neutrino production from this class of sources.

The AGN emission from NGC~1068 reaches a bolometric luminosity of $L_{\rm bol}=10^{44.5}$~erg/s, so if a small fraction of the AGN accretion power is dissipated in non-thermal protons, it can accommodate the neutrino luminosity $L_\nu\simeq 10^{42-43}$~erg/s required to explain the signal observed by IceCube. However, the properties of the produced neutrinos depend on the specific mechanism by which charged particles are accelerated and ultimately transfer some of their energy to neutrinos. A natural candidate region where particle acceleration may happen is the corona of the AGN, a region ranging from a few up to hundreds of gravitational radii~\citep[see e.g.][and references therein]{Cackett2021}, whose existence is postulated to explain the common observation of hard X-rays from non-jetted AGN. The corona provides naturally the required gamma-ray opacity to explain the lack of gamma-ray observations, due to the dense X-ray photon field. This also makes it an efficient neutrino emitter, provided that protons are accelerated within the corona. There are various scenarios which might explain this energization process, including diffusive shock acceleration~\citep{Inoue_2020}, proton re-acceleration in turbulence after a pre-acceleration phase in intermittent reconnection layers~\citep{Mbarek:2023yeq}, or direct acceleration in a macroscopic
reconnection layer~\citep{Fiorillo:2023dts,Karavola:2024uui} (see also, e.g.,~\citet{Khiali:2015tfa}). Another possibility is stochastic acceleration in magnetized turbulence, 
explored by~\cite{Dermer:1995ju} and ~\cite{Murase:2019vdl} 
within the framework of a gyroresonant, weak turbulence model. In~\cite{Fiorillo:2024akm}, we showed that our current understanding of stochastic acceleration -- grounded on first principles particle-in-cell simulations -- enforces the energization process to be non-resonant, and requires strong turbulence, with a magnetic energy density close to the plasma rest mass energy density. \citet{Lemoine24} has also considered a turbulent, non-resonant scenario, with a focus on the effects of acceleration driven by intermittent structures within the turbulent cascade.

Non-thermal production in the coronae of AGN, if confirmed, might serve as a natural explanation for the diffuse neutrino flux observed in the $\sim$10-100~TeV energy range. Indeed, these sources naturally satisfy the constraints on the gamma-ray opacity. Therefore, a reliable estimate of the diffuse neutrino flux from AGN coronae is of primary importance. In fact, the idea that AGN might contribute to the diffuse neutrino flux dates back to earlier times~\citep{Stecker:1991vm}, although many estimates of the diffuse neutrino flux rely on the phenomenological injection of power-law proton fluxes~\citep[e.g.][]{Kalashev:2014vya} rather than the self-consistent modeling of particle acceeleration. \cite{Inoue:2019fil} have estimated the neutrino emission in a scenario of diffusive shock acceleration of protons within a standing shock inside the corona. \cite{Murase:2019vdl} obtained the diffuse neutrino flux in the context of the gyroresonant, weak turbulence model, whereas \cite{Kheirandish_2021} obtained it in the context of a power-law proton injection (this was dubbed as a magnetic reconnection scenario, although we showed in~\cite{Fiorillo:2023dts} that magnetic reconnection generally injects protons with a broken power-law spectrum with properties significantly different from~\cite{Kheirandish_2021}). More recently, \cite{Padovani:2024tgx} obtained the diffuse flux using as a template the point-source flux obtained in~\cite{Murase_2022ApJL} with a power-law proton spectrum, and assuming a simple proportionality between the neutrino and X-ray emission. However, in view of the recent observation~\citep{Fiorillo:2024akm} that stochastic acceleration is non-resonant and requires strongly magnetized environments, it is worth revisiting the systematics associated with the diffuse neutrino flux in the magnetized turbulence scenario.

This is the question we tackle in this work. Our approach is to test whether a consistent scenario -- based on magnetized turbulence -- can accommodate the neutrino emission from the Seyfert galaxies from which IceCube has reported an excess. While these observations are of course preliminary, they allow us to coarsely test whether a unifying coronal scenario can explain their common features. From such a unifying scenario, we then obtain the diffuse neutrino flux and assess its implications.

\section{Physical properties of the corona}\label{sec:coronal_properties}

Our modeling of the source properties follows closely our previous work in~\citet{Fiorillo:2024akm}; we review it here, emphasizing the main differences compared to that work.

We model the corona as a spherical region of size $R$ centered on a black hole of mass $M=M_7\; 10^7\; M_\odot$. The reference length scale is the gravitational radius $r_g=GM/c^2\simeq 1.47\times 10^{12}\;\mathrm{cm}\; M_7$, where $G$ is the gravitational constant and $c$ is the speed of light. The typical values for the coronal size are in the range $R\sim 10-100$~$r_g$; we will later discuss how the results depend on the size $R$.

The corona is permeated by a total (regular and turbulent) magnetic field of total strength $B$ with a stationary turbulent component $\delta B$; the turbulent power is parameterized by $\eta_B=\delta B^2/B^2$. Within the corona, a population of near-thermal particles is established with a number density of leptons $n_e$ and of protons $n_p \simeq n_e$, i.e. we assume no significant pair dominance in this scenario. Only a small fraction of these hadrons are accelerated to very high energies by the magnetized turbulence. The total number density of leptons can be deduced from the Compton opacity of the corona, so that $n_e R \sigma_T=\tau_T\simeq 0.5$; here $\sigma_T$ is the Thomson cross section and $\tau_T$ is the Compton opacity, with a typical value of $0.5$~\citep[e.g.][]{2018MNRAS.480.1819R}. Therefore, the lepton density is
\begin{equation}
    n_e\simeq 2.5\times 10^{10}\;\mathrm{cm}^{-3}\; \frac{20 r_g}{R M_7},
\end{equation}
and the rest-mass energy density of the plasma is
\begin{equation}
    U_{\rm rest}=n_e(m_e+m_p)c^2\simeq 3.9\times 10^7\;\mathrm{erg/cm}^3\;\frac{20 r_g}{R M_7} \, ,
\end{equation}
where $m_e$ and $m_p$ are the masses of the electron and proton, respectively.

The strength of the magnetic field is quantified by
the dimensionless magnetization of the plasma, namely the ratio between magnetic field energy density and rest-mass energy density
\begin{equation} \label{def_sigma}
    \sigma=\frac{B^2}{4\pi U_{\rm rest}} \, .
\end{equation}
We will often use the combination $\sigma_{\rm tur}=\sigma\eta_B$, which defines the ratio between the turbulent energy density and the rest-mass energy density. Our main interest is in the regime $\sigma_{\rm tur}\sim 1$. For $\sigma_{\rm tur}\gg 1$, we rather expect magnetic reconnection, occurring in current sheets formed within the corona, to become the dominant acceleration mechanism, as discussed in~\citet{Fiorillo:2023dts,Karavola:2024uui} (where also pair dominance and a more compact corona was assumed). The total magnetic field is then given by 
\begin{equation}
    B=\sqrt{4\pi\sigma U_{\rm rest}}\simeq 2.2\times 10^4\;\mathrm{G}\;\left(\frac{20 r_g \sigma}{R M_7}\right)^{1/2}.
\end{equation}

Another parameter of interest for our subsequent discussion is the Alfv{\'e}n velocity associated with the turbulent component of the magnetic field, namely $(v_A/c)^2=\sigma_{\rm tur}/(1+\sigma_{\rm tur})\simeq \mathrm{min}(\sigma_{\rm tur},1)$. Finally, the spatial scales of turbulence are characterized by the coherence length $\ell$, which is approximately the scale at which most of the power is injected. We assume that this is a fraction of the coronal size $\ell=\eta R$, with $\eta$ being a dimensionless number less than unity. The corona must also exhibit some mechanism of energy dissipation to maintain the electrons energetic enough to Comptonize photons. Comptonization might be driven either by thermal~\citep[e.g.][]{Yuan:2004qj,Poutanen:2013gra,Fabian:2015wua} or bulk motion~\citep[e.g.][]{Socrates04,Kaufman16,Groselj:2023bgy} of the electrons. The Comptonized X-rays, up to typical energies of $500$~keV, are indeed the primary signature of the coronal activity. Here we do not commit to a specific electron energization scenario, and simply assume that a fraction of the dissipated energy goes to non-thermal protons via magnetized turbulence.

Protons are accelerated by turbulent fluctuations with a typical scale $\lambda\sim \eta R$; since the acceleration mechanism is non-resonant, as shown by particle-in-cell (PIC) simulations~\citep{CS19,2020ApJ...893L...7W,Zhdankin20,2022PhRvD.106b3028B,ComissoFM2024}, as well as by test-particle simulations in magnetohydrodynamic turbulence~\citep{Lynn:2014dya,Kimura:2018clk,Sun:2021ods}, the interaction with these fluctuations is independent of the particle's energy, that differentiated~\citet{Fiorillo:2024akm} from previous works. As discussed there, the typical acceleration timescale for a proton to reach energy $E_p$ is~\citep{CS19}
\begin{equation}
    t_{\rm acc} \simeq \frac{10}{\sigma_{\rm tur}} \frac{\ell}{c} \simeq 10^4\;\mathrm{s}\;\frac{\eta}{\sigma_{\rm tur}}\frac{R M_7}{20 r_g} \, .
\end{equation}
Due to the non-resonant nature of acceleration, the spectrum of the turbulence does not affect the properties of non-thermal protons, which in fact depend purely on the total amount of power within the turbulence at the largest scales, measured by $\sigma_{\rm tur}$.

One point not emphasized in~\citet{Fiorillo:2024akm} is that non-thermal proton production requires Coulomb scattering to be slower than either the injection from the thermal pool or the stochastic acceleration process. The fastest thermalization channel for protons is proton-electron Coulomb scattering, with an energy loss timescale given by~\citep{Murase:2019vdl}
\begin{equation}\label{eq:Coulomb_timescale}
    t_{\rm Coul}\simeq \sqrt{\frac{\pi}{2}}\frac{(\theta_p+\theta_e)^{3/2}}{n_p \sigma_T c \log \Lambda}\frac{m_p}{m_e}\simeq 2.3\times 10^5\;\mathrm{s}\;\frac{RM_7}{20 r_g}\sigma_{\rm tur}^{3/2}.
\end{equation}
Here $\theta_p=k_B T_p/m_p c^2=2r_g/3R\ll 1$, where $T_p$ is the proton temperature and $k_B$ is the Boltzmann constant. If protons are energized by turbulence, we can expect $\theta_p\sim \sigma_{\rm tur}$, assuming the corona is in hydrostatic balance. For electrons, if one assumes thermal Comptonization within the corona, we would expect  $k_B T_e\sim 100$~keV and $\theta_e=k_B T_e/m_e c^2 \sim 0.2$, while for bulk Comptonization, the electron temperature may be much lower. Therefore, to be conservative, we set $\theta_e\ll \theta_p$. Finally, we use $\log\Lambda\simeq 20$. For $\sigma_{\rm tur}\sim 0.1$, we find that $t_{\rm Coul}$ is essentially comparable with the stochastic acceleration timescale, and is much longer than the injection timescale $t_{\rm inj} \simeq \sigma_{\rm tur}/(\eta_{\rm rec} \omega_L)$ \citep{CS18,CS19}, where $\eta_{\rm rec} \simeq 0.1$ is the reconnection rate and $\omega_L$ is the nonrelativistic Larmor frequency. Hence, the injection of relativistic protons is not hindered by thermalization. For higher-energy protons, the thermalization timescale is even longer than that in Eq.~\ref{eq:Coulomb_timescale}, ensuring that proton acceleration is not hindered by Coulomb thermalization.

Due to magnetized turbulence, the typical confinement time of protons is much larger than the light-crossing time of the corona. Following~\citet{Fiorillo:2024akm}, we estimate the escape timescale as
\begin{equation}
    t_\mathrm{esc}\simeq \frac{R}{c}\mathrm{max}\left[1,\frac{R}{\ell }\left(\frac{eB \ell }{E_p}\right)^{1/3}\right], 
\end{equation} 
where $E_p$ indicates the proton energy. In the relevant energy range, this timescale can be estimated as
\begin{equation}
    t_{\rm esc}\simeq 2\times 10^5\;\mathrm{s}\;\left(\frac{E_p}{20\;\mathrm{TeV}}\right)^{-1/3}\eta^{-2/3}\sigma^{1/6}\left(\frac{20r_g}{RM_7}\right)^{7/6}.
\end{equation}

Recently, \cite{Lemoine:2024roa} proposed that an additional escape channel might be more efficient in releasing the non-thermal particles, namely the hydrodynamic diffusion due to turbulent transport. The typical timescale for this process is of the order of $t_{\rm hydro}\simeq R^2/D_{\rm hydro}$, where $D_{\rm hydro}\simeq \ell v_A$, so that one gets $t_{\rm hydro}\simeq R/\eta v_A$. While the most relevant Alfvén velocity may depend on either the turbulent or total magnetic field, this distinction does not materially affect our conclusions, as we consider values of $\eta_B$ that are close to 1. The corresponding timescale can be written as
\begin{equation}\label{eq:hydro_escape}
    t_{\rm hydro}\simeq 10^3\;\mathrm{s}\;\frac{1}{\eta \sqrt{\sigma_{\rm tur}}}\frac{R M_7}{20 r_g}.
\end{equation}

The presence of a strong magnetic field could in principle confine the turbulent motions within a large-scale magnetic configuration, making such an efflux less probable. In this study, we treat this rapid escape as a model-dependent possibility and discuss its impact on the phenomenology in the following section. For now, we note that its effect becomes significant if the hydrodynamic escape timescale is comparable to the acceleration timescale, which occurs for $\sigma_{\rm tur}\lesssim 100 \eta^4$. 

We are here agnostic regarding the fate of the escaping protons, and do not consider their possible neutrino production outside of the turbulent corona. Since all protons ultimately escape, they could in principle interact in the more rarefied photon field over larger distances as they leave the corona, as well as with dust or gas targets. We neglect these potential contributions and focus only on neutrino production within the corona.

In any case, the acceleration of protons to very high energies is not primarily limited by their escape but rather by energy cooling. A subdominant cooling channel is synchrotron radiation over a timescale
\begin{equation}
t_\mathrm{synch}=
    \frac{3m_p^3 c}{2\sigma_T m_e^2 E_p \sigma n_e}\simeq 4.7\times 10^{5}\;\mathrm{s}\;\frac{20\;\mathrm{TeV}}{E_p}\sigma^{-1} \frac{R M_7}{20 r_g}.
\end{equation}
In addition, proton-proton ($pp$) collisions with the bulk of the thermal protons have a characteristic timescale
\begin{equation}
    t_{pp}\simeq(n_p \sigma_{pp}(E_p)\kappa_p c)^{-1},
\end{equation}
where $\kappa_p\simeq 0.5$ is the typical inelasticity of the collision, while $\sigma_{pp}(E_p)$ is the total $pp$ cross section. Unlike \citet{Fiorillo:2024akm}, where the expression from~\citet{1996A&A...309..917A} was used, here we instead adopt the fit valid at all energies from Eq.~(79) of~\citet{Kelner:2006tc}. 

\begin{figure}
    \centering
    \includegraphics[width=0.5\textwidth]{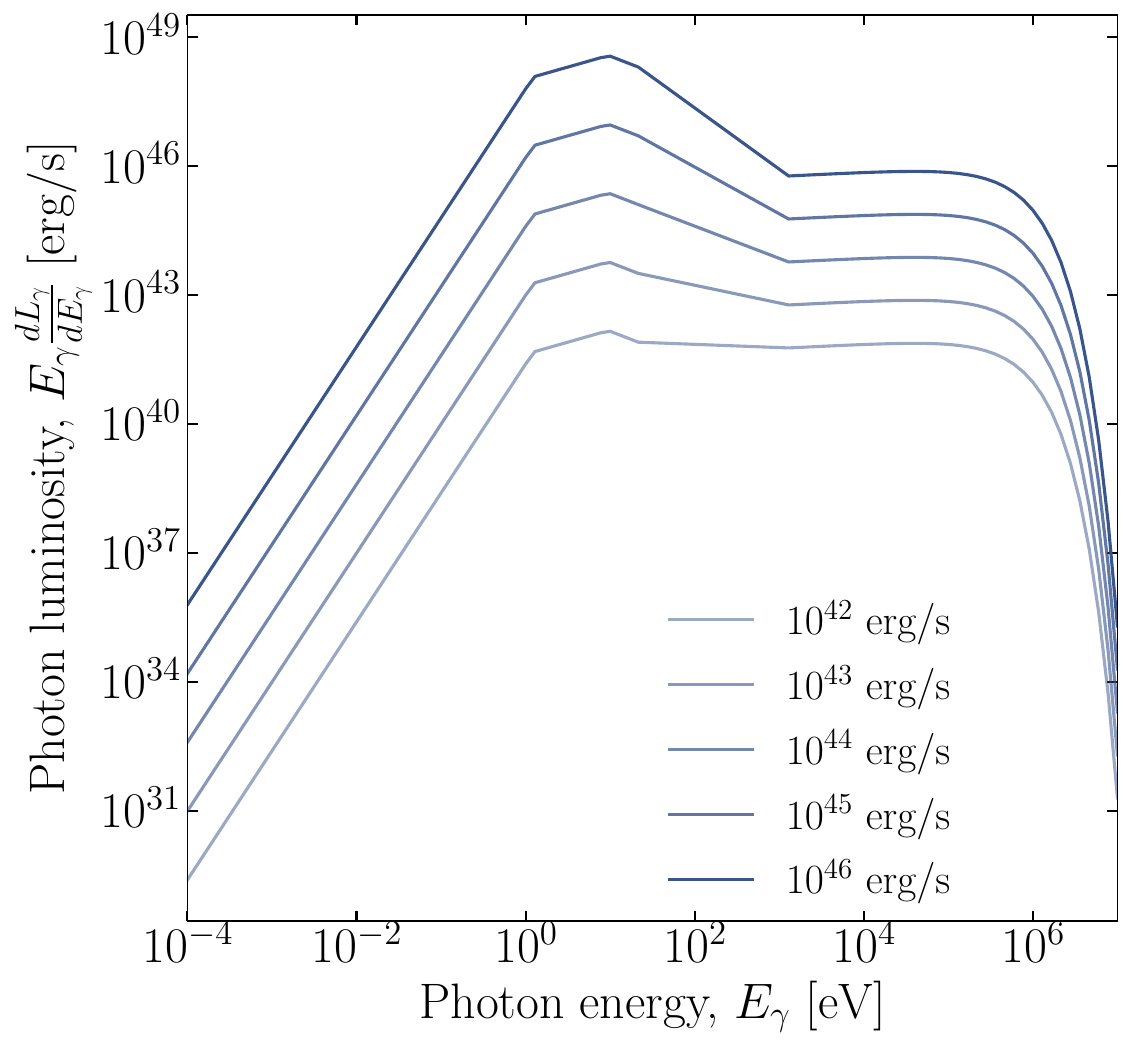}
    \caption{Spectral energy distribution of the AGN photons for varying luminosity. The legend shows the values of $L_X$ for the different curves. The photon spectrum is obtained from the prescription in~\citet{Marconi_2004_SED}.}
    \label{fig:sed}
\end{figure}

Finally, the other relevant proton cooling channels involve scattering on the photon field. Before introducing the relevant timescales, we first discuss the assumed photon field; for our purposes, we are mainly interested in the optical-ultraviolet (OUV) and X-ray field. Here we follow~\citet{Marconi_2004_SED} and model the OUV field as a broken power law, while the X-ray field as a power law with an exponential cutoff at $E_{\gamma,\rm cut}=500$~keV. The indices of each power law are kept fixed to the values reported in~\citet{Marconi_2004_SED}. The normalization of each component is fixed from a single parameter, which we can take to be the X-ray luminosity $L_X$ integrated between $2$~keV and $10$~keV. We denote the total X-ray luminosity, integrated from $1$~keV up to exponential cutoff, as $L_{X,\rm tot}$. Since the spectral shape of the X-rays in the model by~\citet{Marconi_2004_SED} is fixed, the two luminosities are related by $L_{X,\rm tot}\simeq 4.11 L_X$. Our choice of using the luminosity integrated in the 2-10~keV band only is a practical one, since the population properties -- redshift and luminosity evolution, and empirical relations with the central black hole mass -- are conventionally expressed in terms of $L_X$ rather than $L_{X,\rm tot}$. The adopted spectral energy distribution (SED) implies that the X-ray field scales linearly with $L_X$, while the OUV field scales in proportion to $L_X^\alpha$, with $\alpha\simeq 1.6$~\citep{Marconi_2004_SED}. Fig.~\ref{fig:sed} shows the resulting SED for varying AGN luminosity; from this, we extract the photon number density per unit energy
\begin{equation}
    \frac{dN_\gamma}{dV dE_\gamma}=n_\gamma(E_\gamma)=\frac{1}{4\pi R^2 c E_\gamma}\frac{dL_\gamma}{dE_\gamma}.
\end{equation}

The main cooling processes driven by scattering with the photon field are Bethe-Heitler (BH) pair production and inelastic photopion production. For the BH energy losses, driven by the BH process $p\gamma\to p e^+ e^-$, we follow our previous work and use the fits to the cross section and inelasticity from~\citet{1992ApJ...400..181C}. In this context, the dominant target for proton scattering at energies $E_p \sim 20$~TeV are OUV photons with energies on the order of tens of eV. 

Finally, inelastic photopion production, which we denote by $p\gamma$, is modeled following~\cite{Atoyan:2002gu} and~\cite{Dermer:2003zv}. For protons in the tens-of-TeV range, the dominant target photons are now X-rays in the tens-of-keV range. Both the BH and $p\gamma$ energy loss rate expressions remain quite accurate throughout the whole proton energy range, from $E_p \sim 1$~GeV to $E_p \sim 500$~TeV.

\section{High-energy protons and neutrinos in the corona}\label{sec:non_thermal_particles}

In this section, we review the main properties of the non-thermal protons accelerated in the corona, as well as the neutrinos produced via $pp$ and $p\gamma$ interactions. We model the acceleration of the protons with a Fokker-Planck approach in terms of the proton distribution function $f_p$. The differential number of protons is expressed as $dn_p/dE_p= 4\pi p^2 f_p(E_p/c)/c$ for ultra-relativistic protons, where $p$ is the proton momentum and $E_p$ its energy. The transport equation for the protons is modeled as 
\begin{equation}\label{eq:fokker_planck}
    \frac{\partial f_p}{\partial t}=\frac{1}{p^2}\frac{\partial}{\partial p}\left[\frac{p^4}{t_{\rm acc}}\frac{\partial f_p}{\partial p}\right]+\frac{1}{p^2}\frac{\partial}{\partial p}\left[\frac{p^3}{t_{\rm cool}(p)}f_p\right]-\frac{f_p}{t_{\rm esc}}+q_p(p),
\end{equation}
where all the relevant timescales are summarized in Sec.~\ref{sec:coronal_properties}. The cooling timescale is introduced as
\begin{equation}
    t_{\rm cool}^{-1}=t_{pp}^{-1}+t_{p\gamma}^{-1}+t_{\rm BH}^{-1}+t_{\rm synch}^{-1}.
\end{equation}
The injection term $q_p(p)$ is non-vanishing only at low energies. Its specific form is not relevant, since the stationary dynamics at high energies depends only on the balance among stochastic energization, cooling, and escape.

We can obtain the steady-state proton distribution by solving Eq.~\ref{eq:fokker_planck} with the time derivative set to zero. The normalization of the proton spectrum is obtained by relating the amount of energy that the turbulence injects in non-thermal protons per unit time~\citep{Fiorillo:2024akm}
\begin{equation}
    L_p=-4\pi c \frac{4}{3}\pi R^3 \int \frac{p^4}{t_{\rm acc}}\frac{\partial f_p}{\partial p}dp \, , 
\end{equation}
with the rate of magnetic energy dissipation
\begin{equation}
    L_B=\frac{2\pi}{3}\frac{\eta_{\rm rec}}{\eta}\frac{c^3\sigma_{\rm tur}^{3/2}}{(1+\sigma)^{1/2}}(n_p m_p+n_e m_e)R^2 \, ,
\end{equation}
which, for a plasma dominated by the ion mass density, numerically gives
\begin{equation}
    L_B\simeq 2.1\times 10^{44}\frac{\sigma_{\rm tur}^{3/2}}{\eta\;\mathrm{max}(1,\sigma^{1/2})}\frac{R M_7}{20 r_g}\;\mathrm{erg/s} \, .
\end{equation}
This expression is obtained explicitly in~\cite{Fiorillo:2024akm}.
We assume that $L_p=\mathcal{F}_pL_B$, where $\mathcal{F}_p$ serves as a free parameter that governs the normalization of the proton and neutrino spectra. Clearly, it is required that $\mathcal{F}_p<1$.

The magnetic energy dissipation rate allows us to introduce an alternative, and for some purposes more useful, parameterization of the coronal magnetization. In analogy to the fraction of magnetic energy dissipated in non-thermal protons, we can specify the fraction of magnetic energy dissipated in X-rays as
\begin{equation}
    \mathcal{F}_X=\frac{L_{X,\rm tot}}{L_B}.
\end{equation}
If the corona is magnetically powered, we expect $\mathcal{F}_X<1$. Since we are interested in a population study in which $L_X$ varies across the population, it is physically reasonable to consider that $\mathcal{F}_X$, rather than $\sigma_{\rm tur}$, remains roughly of the same order of magnitude across different sources. Therefore, it is convenient to use $\mathcal{F}_X$ as an independent parameter, differently from what we did in~\citet{Fiorillo:2024akm}. Since $L_{X,\rm tot}\simeq 4.11 L_X$, provided that $\sigma\lesssim 1$, we immediately find
\begin{equation}\label{eq:sigma_tur_LX}
    \sigma_{\rm tur}\simeq 0.34 \left(\frac{\eta}{\mathcal{F}_X} \frac{20 r_g}{R M_7} \frac{L_X}{10^{43}\;\mathrm{erg/s}}\right)^{2/3}.
\end{equation}

Once we have obtained the steady-state proton distribution, we can proceed to calculate the spectrum of the neutrinos produced through $pp$ and $p\gamma$ interactions. We follow here the same procedure as in~\citet{Fiorillo:2024akm}, which in turn uses the fit functions from~\citet{Kelner:2006tc}
 and~\citet{Kelner:2008ke}. We neglect the possible cooling of secondary muons and pions in the coronal magnetic field. As we will see, the peak neutrino energies we find justify this choice. 

 \begin{figure*}
     \includegraphics[width=\textwidth]{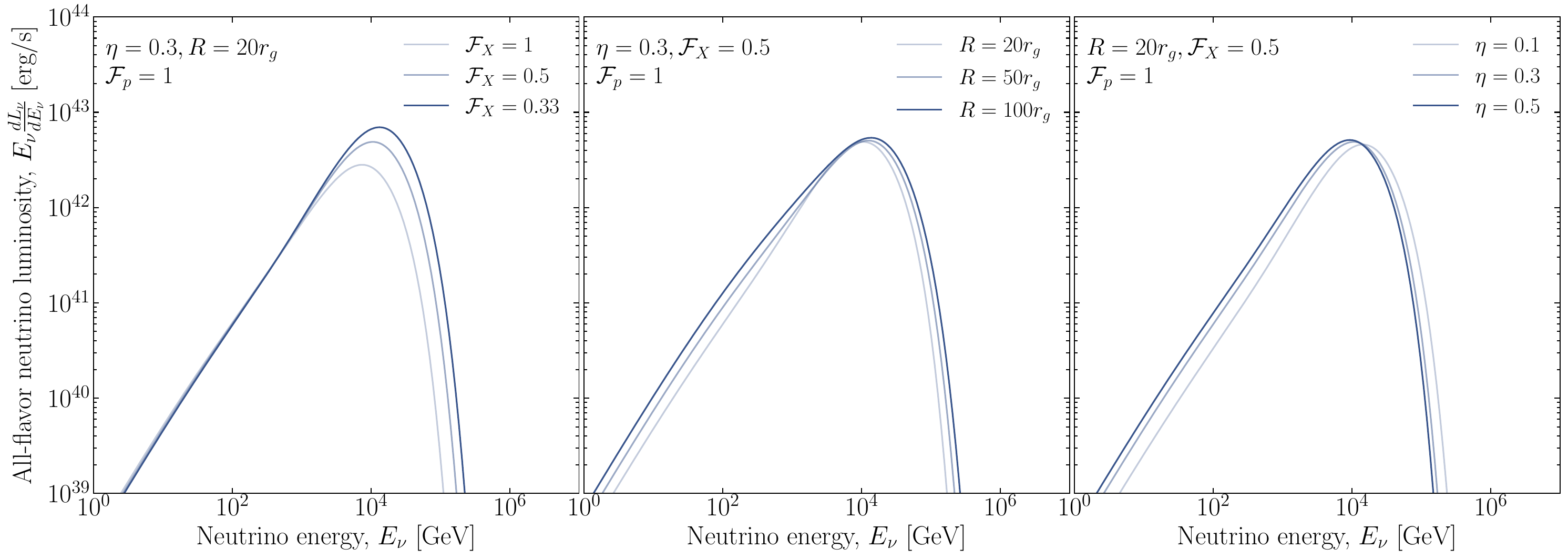}
     \caption{Impact of coronal parameters on the neutrino production from NGC~1068. In each panel, we independently vary each of the parameters $\mathcal{F}_X$, $R$, and $\eta$. The observed flux has been converted to intrinsic source luminosity using a luminosity distance for the source of $d_L=10.1$~Mpc~\citep{Padovani:2024ibi}.}
     \label{fig:ngc1068}
 \end{figure*}

Let us summarize our model parameters that determine the neutrino production in the corona: these include parameters related to the AGN properties, namely the X-ray luminosity $L_X$ (which, within the model of~\citet{Marconi_2004_SED}, determines the whole SED of the AGN) and the black hole mass $M$, as well as parameters related to the coronal properties, namely the coronal radius $R$, the coherence length expressed in terms of the dimensionless parameter $\eta$, and the energy fractions $\mathcal{F}_X$ and $\mathcal{F}_p$. 
 
 To explore the role of these parameters in determining the properties of the produced neutrinos, we consider the typical parameters of NGC~1068 (see Table~\ref{tab:astrophysical_params} for the baseline values assumed for $L_X$ and $M$). Fig.~\ref{fig:ngc1068} shows the neutrino spectrum that we obtain for varying values of $\mathcal{F}_X$, $R$, and $\eta$. In all cases, for illustration, we set $\mathcal{F}_p=1$, recognizing $\mathcal{F}_p$ serves solely as a normalization factor for the produced neutrino flux, and therefore the results for different values can simply be scaled down proportionally.

 We find that varying the parameters within their allowed range does not have a dramatic impact on the spectral shape. The most impactful parameter is $\mathcal{F}_X$; when this is reduced, the magnetic field energy density and the magnetization $\sigma_{\rm tur}$ increase. Since $\mathcal{F}_p$ is fixed, the proton luminosity, as well as the peak neutrino luminosity, increase in proportion to $\mathcal{F}_X^{-1}$. In addition, decreasing $\mathcal{F}_X$ leads to an increase in $\sigma_{\rm turb}$ and in the acceleration rate, and therefore to a larger value for the peak energy of the produced neutrinos. The impact of $R$ and $\eta$ is more limited. Increasing $R$ gives a less compact corona, and therefore to slightly less pronounced radiative losses, with the spectrum peaking at larger energies. However, because the Bethe-Heitler loss timescale drops very rapidly with energy in the range in which it becomes comparable with the acceleration timescale, the effect is very marginal. The same happens by increasing the coherence parameter $\eta$, which leads to a reduced acceleration rate and in turn to a slight reduction in the peak energy for the neutrinos. In all cases, the neutrino spectrum at low energies is dominated by $pp$ production and has a characteristic $dL_\nu/dE_\nu\propto E_\nu$ scaling inherited from the parent proton spectrum, as was shown in~\citet{Fiorillo:2024akm}.

Under the baseline values adopted for $L_X$ and $M$, the peak neutrino energy for NGC~1068 is found to exceed the range suggested by the IceCube signal. The very mild dependence of this result on the coronal parameters might suggest a tension of the model with the observations. However, we emphasize that there are several factors, both observational and theoretical, that could reduce this discrepancy. From the observational side, the uncertainty band in the flux identified by IceCube strongly depends on the assumed spectral shape for the reconstruction analysis -- for instance, employing the alternative model used in~\citet{IceCube24_Seyfert} results in a significantly higher peak energy than the flux we report here -- and even the two different studies from the IceCube collaboration~\citep{IceCube-NGC1068,IceCube24_Seyfert} yield noticeably different spectra. From the theoretical side, our prediction for the neutrino spectral shape depends sensitively on the black hole mass $M$ and X-ray luminosity $L_X$ -- a smaller mass, with a more compact corona, and a higher X-ray luminosity both produce larger photohadronic cooling, which reduces the neutrino peak energy. It also strongly depends on the assumed spectral shape of the SED, which determines the precise energy at which Bethe-Heitler losses set in and determine the peak of the neutrino spectrum. We rely on the specific model by~\citet{Marconi_2004_SED}, but a larger photon flux in the energy band between 10--100~eV, the dominant target for Bethe-Heitler scattering off protons with energies between 10--100~TeV, could significantly increase the photohadronic losses in the energy range identified by IceCube and lower the neutrino peak energy. Given the relatively coarse experimental uncertainties at present, we do not attempt a more detailed modeling of the SED at this stage. Notice also that the typical peak neutrino energy that we find is at most 10~TeV. This justifies our choice of neglecting the synchrotron cooling of secondary muons and pions. The typical muon energy required to produce these neutrinos is about 30~TeV. In a magnetic field $B=2.2\times 10^4$~G, as is the case for NGC~1068 (using the values in Table~\ref{tab:astrophysical_params}), the critical energy above which the muon cooling time is shorter than the decay time is $E_{c,\mu}\sim 60$~TeV (see, e.g., Eq.~(B.3) of~\citet{Fiorillo:2021hty}). For pions, the critical energy is even higher. 

 Finally, we address the potential feedback of the proton energization on the turbulent cascade itself, which in the context of NGC~1068 has been brought up by~\citet{Lemoine24,Lemoine:2024roa}. As a measure of the dynamical impact of non-thermal protons, we can compare the proton energy density $U_p=\int dE_p E_p dn_p/dE_p$ with the turbulent magnetic energy density $U_B=\sigma_{\rm tur} U_{\rm rest}/2$. For the baseline values $\eta=0.3$, $R=20r_g$, $\mathcal{F}_X=0.5$, we find that
 \begin{equation}
     \frac{U_p}{U_B}\simeq 0.32 \mathcal{F}_p.
 \end{equation}
 From Fig.~\ref{fig:ngc1068}, we see that a value of $\mathcal{F}_p\sim 0.1$ is favored by the IceCube measurements. This suggests that non-thermal protons have a relatively minor dynamical impact on the development of magnetized turbulence. While the precise value of $U_p/U_B$ depends on the shape of the proton energy distribution, in particular due to the pile-up effect that we noticed in~\citet{Fiorillo:2024akm}, its qualitative behavior is mostly independent of the specific energy losses. In fact, $U_p$ can be approximately obtained assuming that the energy injected in high-energy protons per unit time, $L_p=\mathcal{F}_p L_B$, is dissipated within a typical timescale $t_{\rm acc}$, so that
 \begin{equation}
     \frac{U_p}{U_B}\sim\frac{\mathcal{F}_p L_B t_{\rm acc}}{\frac{4}{3}\pi R^3 U_B}\simeq\frac{\mathcal{F}_p}{\sqrt{\sigma_{\rm tur}}}.
 \end{equation}
 We have verified that this approximate relation correctly captures the scaling with $\sigma_{\rm tur}$ that we find from the numerical solution of the Fokker-Planck equation, and in fact it holds to a very good approximation even quantitatively when multiplied by a factor $0.25$.

 \section{Neutrino emission from a population of AGN coronae}\label{sec:results}

  \begin{table} 
    \centering
    \renewcommand{\arraystretch}{1.2}
    \begin{tabular}{lccc}
        \toprule
        Parameter & NGC~1068 & NGC~4151 & CGCG~420-015 \\
        \midrule
        $L_X$ [$10^{43}$ erg/s] & 3 & 0.8 & 7 \\
        $M$ [$10^7\;M_\odot$] & 0.67  & 4.6 & 20 \\
        $\sigma_{\rm tur}$ &  0.66   &   0.075  & 0.12 \\       
        \bottomrule
    \end{tabular}
    \caption{AGN parameters adopted for each of the Seyfert galaxies to which we apply our model (X-ray luminosities have been taken from \citet{Padovani:2024ibi,Gianolli_2023,Tanimoto2018} respectively, while black hole masses are taken from~\citet{Padovani:2024ibi, Bentz:2006ks, Koss:2017ilk} respectively). The value of $\sigma_{\rm tur}$ is obtained from Eq.~\ref{eq:sigma_tur_LX} using the benchmark choice $\eta=0.3$, $R=20r_g$, $\mathcal{F}_X=0.5$.
    }
    \label{tab:astrophysical_params}
\end{table}

 \begin{figure*}
     \includegraphics[width=\textwidth]{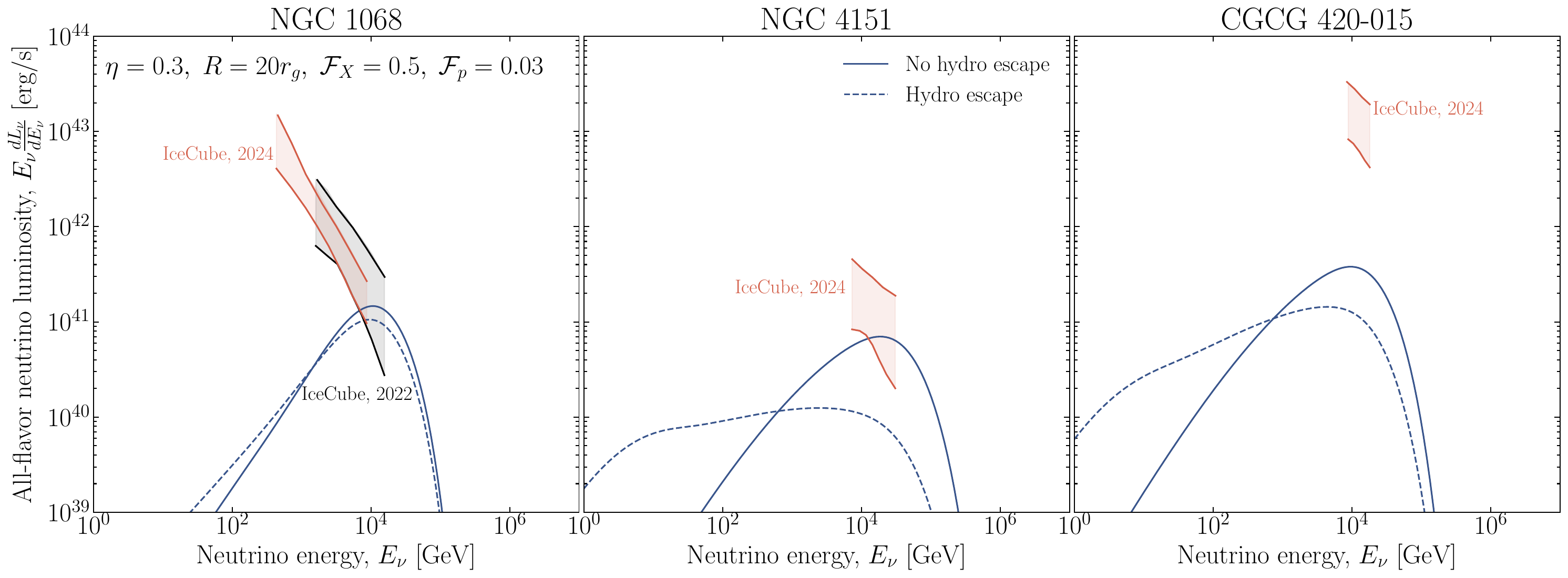}
     \caption{Neutrino production from Seyfert galaxies from which IceCube has reported a neutrino excess~\citep{IceCube24_Seyfert}. We show the neutrino spectra for our baseline choice of parameters reported in the figure. The AGN luminosity and black hole mass adopted for each galaxy are summarized in Table~\ref{tab:astrophysical_params}. We also show results for the model which includes the hydrodynamical turbulent escape, as discussed in the main text.  The flux bands are extracted from the recent IceCube results~\citep{IceCube24_Seyfert}, while for NGC~1068 we also show the results from the 2022 study~\citep{IceCube-NGC1068}.}
     \label{fig:seyfert}
 \end{figure*}

 In order to extend our analysis to a population study, beyond the application to NGC~1068 alone, we first consider the three Seyfert galaxies from which IceCube has recently reported the detection of a neutrino excess~\citep{IceCube24_Seyfert}. In applying our model, we fix the benchmark parameters for the corona identified at the end of Sec.~\ref{sec:non_thermal_particles}. As discussed in Sec.~\ref{sec:non_thermal_particles}, changing these parameters has only a moderate impact on the spectrum of the produced neutrinos. Regarding the AGN parameters, we summarize the baseline values used in this work in Table~\ref{tab:astrophysical_params}.

 Fig.~\ref{fig:seyfert} shows the neutrino spectrum we obtain for each of the Seyfert galaxies analyzed. We only show results for $\mathcal{F}_p=0.03$, with the understanding that these can be scaled up or down directly to obtain the results for different values of $\mathcal{F}_p$. For all galaxies, our model leads to predictions of peak neutrino energy in rough agreement with the energy range identified by IceCube. For NGC~1068 and NGC~4151, a value $\mathcal{F}_p\sim 0.01-0.1$ is required to match the normalization of the IceCube excess at the relevant energy range. Instead, for CGCG~420-015, a value $\mathcal{F}_p\sim 1$ seems to be required to match the large excess reported by IceCube. This aligns with findings by~\citet{Karavola:2024uui}, who also suggest particularly efficient neutrino production from this galaxy. Given the current uncertainties in these excesses, we do not interpret this as a significant tension of the model.

 In Fig.~\ref{fig:seyfert}, we also show how the neutrino spectrum changes when introducing the hydrodynamic turbulent escape, as described by the additional escape timescale in Eq.~\ref{eq:hydro_escape}. For NGC~4151 and CGCG~420-015, introducing this additional escape channel significantly reduces the flux of emitted neutrinos and softens the spectrum at low energies. This is due to the large fraction of particles that manages to escape from the turbulent region before depositing their energy in $p\gamma$ interactions. However, this statement should be regarded as model-dependent, as in reality such non-thermal particles may still interact with the radiation field in the regions surrounding the corona; the neutrino production depends now on the details of the region in which the protons escape. At lower energies, the neutrino spectrum is softened due to the competition between escape and acceleration; as discussed in~\citet{Fiorillo:2024akm}, in this energy range we expect the neutrino energy spectrum to follow $dL_\nu/dE_\nu\propto E_\nu^{1-t_{\rm acc}/t_{\rm hydro}}$. Notice that for NGC~1068 the consequences of the more rapid escape are much more limited; this is due to the larger magnetization (see Table~\ref{tab:astrophysical_params}), which causes the acceleration timescale to be much closer to the escape timescale.

 \begin{figure*}
     \includegraphics[width=0.431\textwidth]{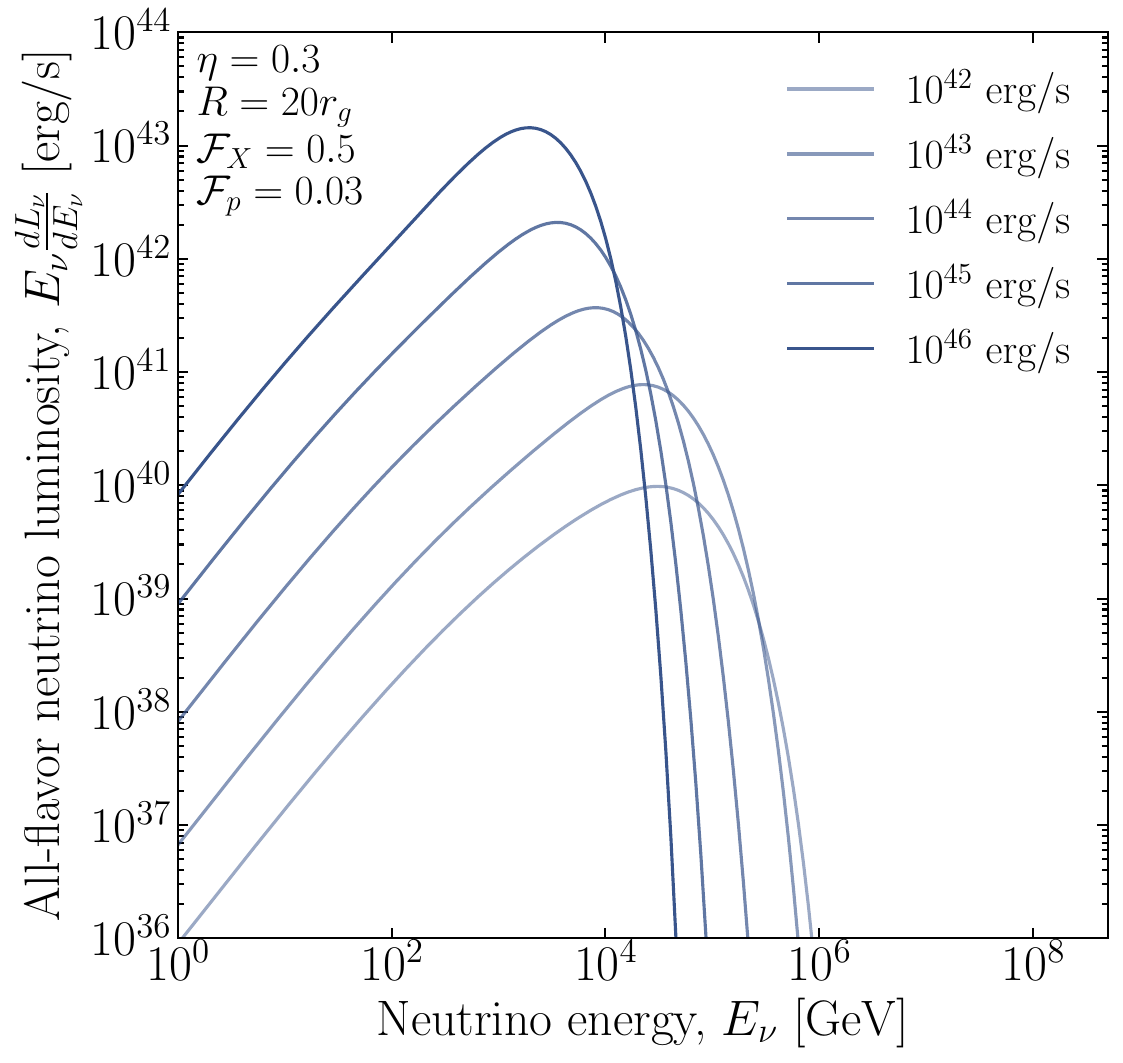}
     \includegraphics[width=0.47\textwidth]{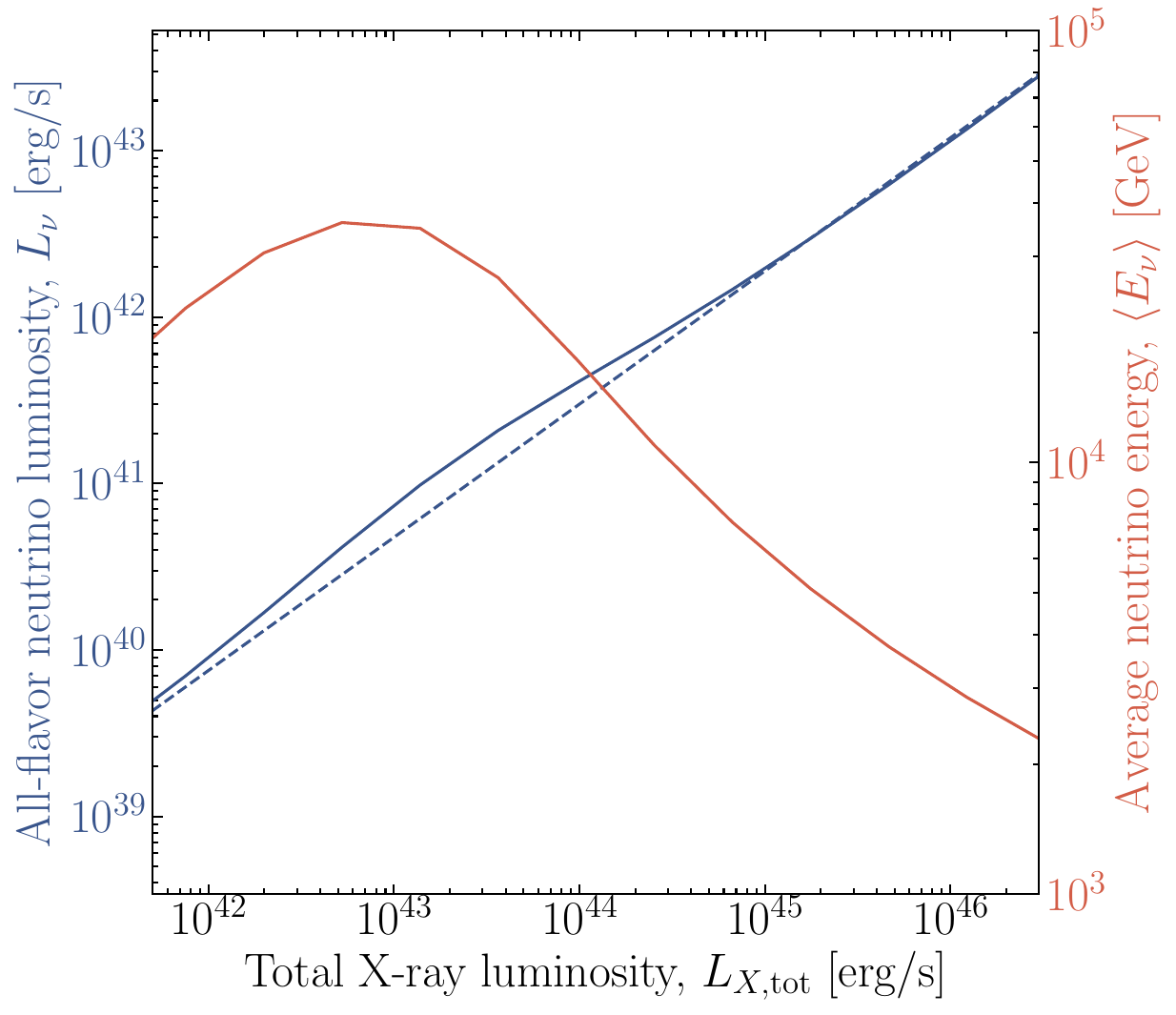}
     \caption{\textit{(Left)} Neutrino flux for varying X-ray luminosities; in the legend, we report the 2-10~keV luminosity $L_X$. The chosen values are the same as for the AGN SED in Fig.~\ref{fig:sed}. \textit{(Right)} Neutrino luminosity and characteristic peak neutrino energy as a function of the total X-ray luminosity $L_{X,\rm tot}$. We choose to use $L_{X,\rm tot}$, rather than $L_X$, to highlight more clearly the fraction of the total X-ray power that is emitted in neutrinos. We show with a dashed line the approximate fit in Eq.~\ref{eq:fit_luminosity} for the $L_\nu-L_{X,\rm tot}$ relation.}
     \label{fig:evolution}
 \end{figure*}

 Building on the reasonable agreement with the neutrino excesses reported by IceCube, we can now proceed to compute the luminosity-dependent neutrino emission from Seyfert galaxies. We adopt a common scenario, based on a standard-candle approximation in which all galaxies have the same value of $\eta$, $\mathcal{F}_X$, $R/r_g$ used for our previous results. In addition, we choose a constant $\mathcal{F}_p=0.03$, representative of the range that leads to a reasonable agreement with the IceCube excess for NGC~1068 and NGC~4151. Finally, to obtain the luminosity-dependent neutrino emission, we also need to specify how the black hole mass evolves with the AGN luminosity. Here we adopt the simplifying assumption of a one-to-one relation between the two quantities, following Fig.~11 of~\citet{Mayers:2018hau} which relates $M$ to $L_X$ in the 2-10~keV band
 \begin{equation}
     M=3\times10^7\; M_\odot\;\left(\frac{L_X}{2\times 10^{43}\;\mathrm{erg/s}}\right)^{0.58}.
 \end{equation}
Notice that our relations are only mildly dependent on the particular relation we assume; we will show that the diffuse neutrino flux is dominated by Seyfert galaxies within the same luminosity range as NGC~1068, which have therefore similar typical properties.

 With this relation, and the standard-candle approximation, the neutrino emissivity depends only on a single parameter, namely the X-ray luminosity $L_X$. For the same values of $L_X$ for which we showed the SED from the AGN in Fig.~\ref{fig:sed}, we now show the neutrino emission for varying $L_X$ in Fig.~\ref{fig:evolution} (left panel). As the X-ray luminosity increases, the neutrino emission increases in intensity, but not by a trivial scaling of the normalization. At low X-ray luminosities, increasing $L_X$ leads to a higher acceleration rate (since for a fixed $\mathcal{F}_X$ larger luminosities lead to larger magnetizations and more rapid acceleration), and therefore to a higher peak neutrino energy. However, when the luminosity rises above $L_X\gtrsim 10^{42}$~erg/s, increasing it further leads to stronger photohadronic losses which reduce, rather than increase, the peak neutrino energy. This trend is visible also in the right panel of Fig.~\ref{fig:evolution}, where we show the characteristic peak neutrino energy defined as
 \begin{equation}
     \langle E_\nu\rangle=\frac{\int \frac{dL_\nu}{dE_\nu}E_\nu dE_\nu}{\int \frac{dL_\nu}{dE_\nu}dE_\nu}.
 \end{equation}
This average energy peaks in the range of a few tens of TeV when the total X-ray luminosity $L_{X,\rm tot}\sim 10^{43}$~erg/s. Generally, the peak energy never exceeds a few tens of TeV, so that a turbulent corona is unable to explain neutrino production, either diffuse or from specific AGN point sources, above tens of TeV. This rules out the possibility that magnetized turbulence could explain neutrino production in the coronal region of blazars such as TXS~0506+056, from which IceCube has reported the detection of neutrinos up to PeV; see also the dedicated discussion in~\cite{Fiorillo:2025cgm}.
We also show the integrated neutrino luminosity $L_\nu$ as a function of $L_{X,\rm tot}$; we find that the behavior is generally not linear, and an approximate fit to the trend shown in Fig.~\ref{fig:evolution} is 
\begin{equation}\label{eq:fit_luminosity}
    L_\nu\simeq 2\times 10^{41}\;\mathrm{erg/s}\;\left(\frac{L_{X,\rm tot}}{10^{44}\;\mathrm{erg/s}}\right)^{0.8},
\end{equation}
a sub-linear scaling.
We show this approximate fit as a dashed line in Fig.~\ref{fig:evolution}. We conclude that the neutrino production throughout the AGN sequence cannot be trivially represented as a linear scaling in normalization of the neutrino spectrum with $L_{X}$, as adopted, e.g., in~\citet{Padovani2024}, and it must account for the evolution in the spectral shape with X-ray luminosity.

\begin{figure}
    \centering
    \includegraphics[width=0.5\textwidth]{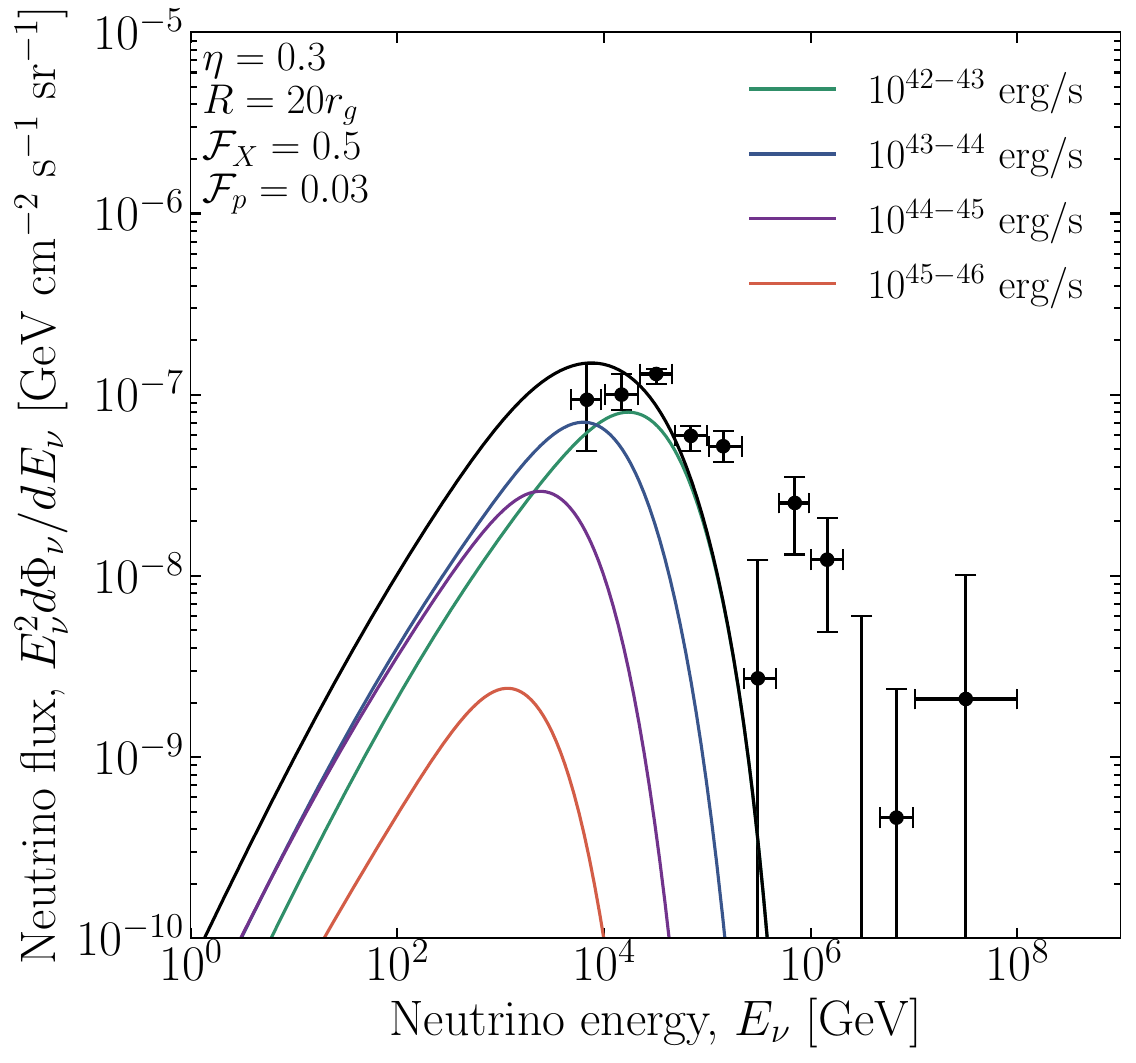}
    \caption{Predicted diffuse neutrino flux (black) from the AGN population. We show the reported flux data points from the IceCube collaboration~\citep{Naab:2023xcz}. We also show the separate contributions from different bands of $L_X$ with colored curves.}
    \label{fig:diffuse}
\end{figure}

We can finally estimate the diffuse neutrino flux from the cosmological population of Seyfert galaxies. This requires us to parameterize the luminosity and redshift evolution of AGN. We adopt the prescription for the luminosity-dependent density evolution (LDDE) of~\citet{Ueda:2014tma}, introducing $d\Phi/d\log_{10}L_X$ (measured in Mpc$^{-3}$ dex$^{-1}$) as the number of sources per comoving volume per decade of luminosity. Notice that~\citet{Ueda:2014tma} provides the distribution in terms of the 2-10~keV luminosity $L_X$ measured in the source rest frame. The number of neutrinos emitted per unit time and energy in the rest frame of the AGN is $dN_\nu/dE_\nu dt=E_\nu^{-1}dL_\nu/dE_\nu$,
and the diffuse neutrino flux at Earth is
\begin{equation}
    \frac{d\Phi_\nu}{dE_\nu}=\frac{c}{4\pi}\int_0^{+\infty}\frac{dz}{H(z)} \int dL_X \frac{d\Phi}{dL_X} \frac{dN_\nu}{dE_\nu dt}\left[E_\nu (1+z)\right].
\end{equation}

The resulting diffuse flux is shown in Fig.~\ref{fig:diffuse}, where we compare it with the IceCube measurements from a combined fit of the different IceCube data samples. The AGN contribution can explain the whole of the neutrino production in the $\sim 1-100$~TeV energy range, while it steeply decreases at higher energies. The diffuse neutrino production is dominated by local AGN ($z\ll 1$) with a typical luminosity in the range $10^{42}~\mathrm{erg/s}\lesssim L_X\lesssim 10^{44}\;\mathrm{erg/s}$. To confirm this, we also show separately the contribution to the diffuse flux from separate ranges of $L_X$. Notice that all of the Seyfert galaxies from which IceCube has reported a neutrino excess lie in these ranges of $L_X$. We do not show results for $L_X\lesssim 10^{42}$~erg/s, to avoid extrapolating the model used for the SED~\citep{Marconi_2004_SED} in regions where very few AGN are observed; we have explicitly verified that the overall contribution to the diffuse flux from this low-luminosity range is negligible. Finally, notice that in order to match with the magnitude of the diffuse flux, we chose a relatively small proton power, with $\mathcal{F}_p=0.03$. For NGC~1068 and NGC~4151, such a small proton power is still roughly consistent with the IceCube observations (see Fig.~\ref{fig:seyfert}), while, as already discussed, CGCG~420-015 requires $\mathcal{F}_p\sim 1$ to reproduce the observed excess. Due to the currently large uncertainties on the specific properties of the neutrino excesses, we do not interpret this as a significant tension, especially in light of our standard-candle assumption. 

\section{Summary and discussion}\label{sec:discussion}

Using the information from individual neutrino point sources to understand the properties of the diffuse neutrino flux detected by IceCube is a timely and crucial task. In this work, we have focused on the case of AGN coronae, within the context of the strongly magnetized turbulence model proposed in~\citet{Fiorillo:2024akm}. Our general strategy has been to first test the dependence of the model predictions on the coronal properties (size, coherence length, magnetization), which cannot be generally constrained from electromagnetic measurements at present. We have shown that the predictions are relatively robust to variations in these parameters. Due to the relevance of photohadronic cooling, which sets the maximum energy scale at which protons can be accelerated, a much more sensitive dependence is expected on the precise shape of the electromagnetic SED of the AGN; here we have adopted the model prediction of~\citet{Marconi_2004_SED}, but determining the impact of a different choice would be an important step for a more precise approach. 
In addition, our inferred magnetic fields are about two orders of magnitude higher than the ones inferred by the recent interpretation of the radio signal between 100 and 1000 GHz from a few AGN as being of coronal origin~\citep{Inoue:2018kbv,Michiyama:2023okx,delPalacio:2025ibd}. Such inference is based on specific assumptions on the non-thermal electron properties within the corona, e.g. a given ratio between non-thermal particle and magnetic field energy density, and between non-thermal and thermal particle density, isotropy of the accelerated leptons, absence of secondary non-thermal lepton injection. Relaxing the simplifying assumptions of these one-zone models would be crucial to obtain definite conclusions on the coronal properties, as the energy density of non-thermal leptons is degenerate with the magnetic field strength as a parameter. In fact, magnetized turbulence would also likely lead to pitch angle anisotropy for the non-thermal leptons~\citep{CS19,CS21,ComissoApJL2020}, while high-energy proton and gamma-rays would inject secondary leptons via Bethe-Heitler, $\gamma\gamma$, and $p\gamma$ interactions. For all of these reasons, we plan to discuss as a follow-up work a comprehensive analysis including both millimeter band and neutrino observations, to test their potential connection.

We have subsequently tested the predictions of the strongly magnetized turbulent model for the Seyfert galaxies from which IceCube has recently reported a neutrino excess. We find a generally consistent picture, suggesting that protons may take 1--10$\%$ of the magnetically dissipated energy in the turbulent corona. The case of CGCG~420-015, as also discussed in other works~\citep{Karavola:2024uui}, requires a somewhat more efficient neutrino production, with $\mathcal{F}_p\sim 1$. We should stress however that the precise properties of the excesses reported by IceCube are still preliminary, and dependent on the assumed spectral shape in the IceCube analysis.

Based on the parameters inferred by these excesses, we have obtained the luminosity-dependent neutrino emission from AGN. This is a key result of this work, showing that simplifying assumptions, such as that the neutrino spectrum simply scales in normalization linearly with the X-ray luminosity, do not generally hold in this context. The reason is that in the strongly magnetized turbulence scenario, increasing the X-ray luminosity leads to an increase both in the acceleration rate, due to the larger magnetization of brighter coronae, and in the photohadronic cooling. The competition between the two factors leads to a non-trivial behavior, with the neutrino luminosity increasing approximately as $L_\nu\propto L_X^{0.8}$, and the peak neutrino energy increasing up to a maximum of $10^4$~TeV at $L_X\sim 10^{42}$~erg/s and decreasing for larger luminosities due to the strong cooling. These conclusions crucially depend on our assumption that, as $L_X$ increases, the fraction $\mathcal{F}_X$ remains the same. Intuitively, this is motivated by the general idea that X-rays, produced by leptonic Comptonization, are ultimately powered by the magnetic field. However, this should be regarded as a model assumption. Our results also substantially differ from those of~\citet{Ambrosone:2024zrf}, which adopted a phenomenological leaky-box model with a fixed injection spectrum, not based on any specific acceleration mechanism. For turbulent coronae, the situation is quite different, since acceleration happens within the same region as photohadronic losses, so the resulting neutrino luminosity and spectral shape have a quite different evolution with the AGN luminosity. More generally, the AGN neutrino sequence we have constructed relies on a standard-candle approximation, in which the dimensionless properties of the corona -- the relative size $R/r_g$, the relative coherence length $\eta$, and the fraction of magnetic energy dissipated in X-rays $\mathcal{F}_X$ and non-thermal protons $\mathcal{F}_p$ -- are the same for all AGN. Furthermore, it relies on a specific choice of relation between $M$ and $L_X$, which is inferred from empirical correlations~\citep{Mayers:2018hau}. A natural step beyond our work is therefore to test the impact of lifting some of these assumptions.

Within the scenario that we have constructed, we have finally tested the properties of the diffuse neutrino flux from the AGN population. In order not to overshoot significantly the diffuse flux observed by IceCube in the 10--100~TeV range, we find that a relatively small fraction of the magnetic energy should be dissipated in non-thermal protons, of the order of a few percent. This is still roughly consistent with the properties of the neutrino excesses reported by IceCube from individual Seyfert galaxies. The diffuse flux contribution due to AGN drops rapidly above a few tens of TeV, indicating that, within the corona model based on strongly magnetized turbulence, neutrinos above 100~TeV cannot be explained by AGN cores. This is ultimately due to the rapid Bethe-Heitler cooling, that limits acceleration to higher energies. This conclusion depends ultimately on the assumed acceleration mechanism; for example, reconnection-based scenarios~\citep{Fiorillo:2023dts,Karavola:2024uui} may in principle explain neutrino production up to higher energies, even though for the case of NGC~1068 the peak neutrino energy is favored to be in the range of 1--10~TeV.

Overall, our work suggests that a consistency picture between the neutrino excesses from individual Seyfert galaxies and the diffuse neutrino flux in the $\sim 1-100$~TeV energy range is available within the strongly magnetized turbulent corona model. Such a consistency picture depends however on the relatively coarse precision of the reconstructed spectral shape and normalization of the neutrino excess from the individual Seyfert galaxies, and it may be challenged by more precise reconstructions of these properties. This might hint at a violation of our assumptions, especially the standard-candle assumption according to which the intrinsic coronal properties remain identical for all AGN coronae. Such an assumption is adopted here only as a starting point, given the limited amount of observational data available at present, and identifying potential observables that might put it to the test is an important step to constrain this theoretical scenario further.

\section*{Acknowledgements} 

We thank Mahmoud Alawashra for useful comments on an early draft of this work. D.F.G.F. is supported by the Alexander von Humboldt Foundation (Germany).
L.C is supported by the NSF grant PHY-2308944 and the NASA ATP grant 80NSSC22K0667.
EP acknowledges support from INAF through ``Assegni di ricerca per progetti di ricerca relativi a CTA e precursori'' and from the Agence Nationale de la Recherche (grant ANR-21-CE31-0028).
M.P. acknowledges support from the Hellenic Foundation for Research and Innovation (H.F.R.I.) under the ``2nd call for H.F.R.I. Research Projects to support Faculty members and Researchers'' through the project UNTRAPHOB (ID 3013). 
L.S. acknowledges support from DoE Early Career Award DE-SC0023015, NASA ATP 80NSSC24K1238, NASA ATP 80NSSC24K1826, and NSF AST-2307202. This work was supported by a grant from the Simons Foundation (MP-SCMPS-00001470) to L.S. and facilitated by Multimessenger Plasma Physics Center (MPPC), grant NSF PHY-2206609 to L.S.

\appendix
\section{Impact of hydrodynamical escape}

\begin{figure*}
    \includegraphics[width=\textwidth]{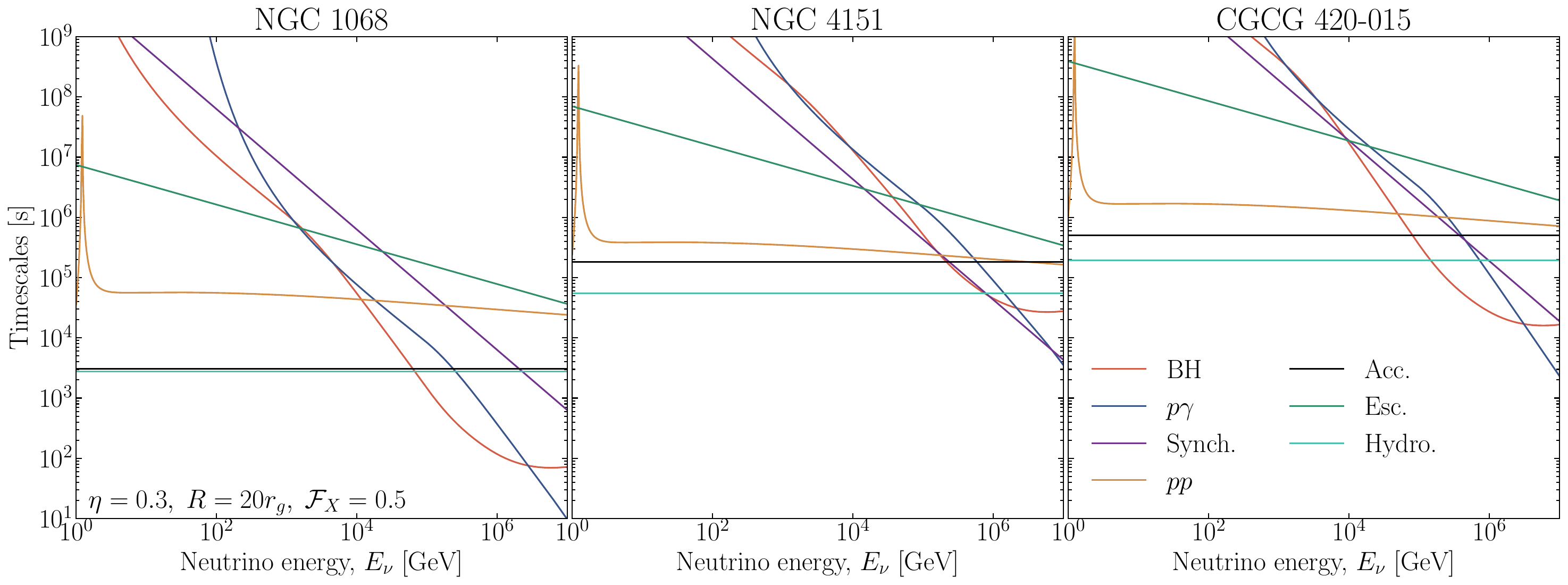}
    \caption{Energy-dependent timescales for the benchmark scenarios presented in Fig.~\ref{fig:seyfert}; we include Bethe-Heitler (BH), acceleration (Acc.), proton-photon ($p\gamma$), diffusive escape (Esc.), synchrotron (Synch.), hydrodynamical escape (Hydro.), and proton-proton ($pp$).}\label{fig:timescales}
\end{figure*}

In this section, we elaborate in more detail on the effect of the hydrodynamical escape. In Fig.~\ref{fig:seyfert}, we have seen that this additional escape channel tends to soften the neutrino spectrum and lower its intensity, due to a higher fraction of protons leaving the acceleration region. As discussed before, such statements must be regarded as model-dependent, since in reality even the escaped protons might interact with additional target outside of the acceleration region and produce further neutrinos. Fig.~\ref{fig:timescales} shows the timescales for acceleration, escape, and energy loss for the protons for all three Seyfert galaxies. The hydrodynamical escape timescale is comparable with the acceleration timescale for NGC~1068, whereas it is shorter by a factor 2-3 for the other two galaxies. This is essentially the reason for the much more significant impact of the hydrodynamical escape on NGC~4151 and CGCG~420-015 than on NGC~1068; the latter is more strongly magnetized, and therefore its acceleration is more rapid.

\begin{figure}
    \centering
    \includegraphics[width=0.5\textwidth]{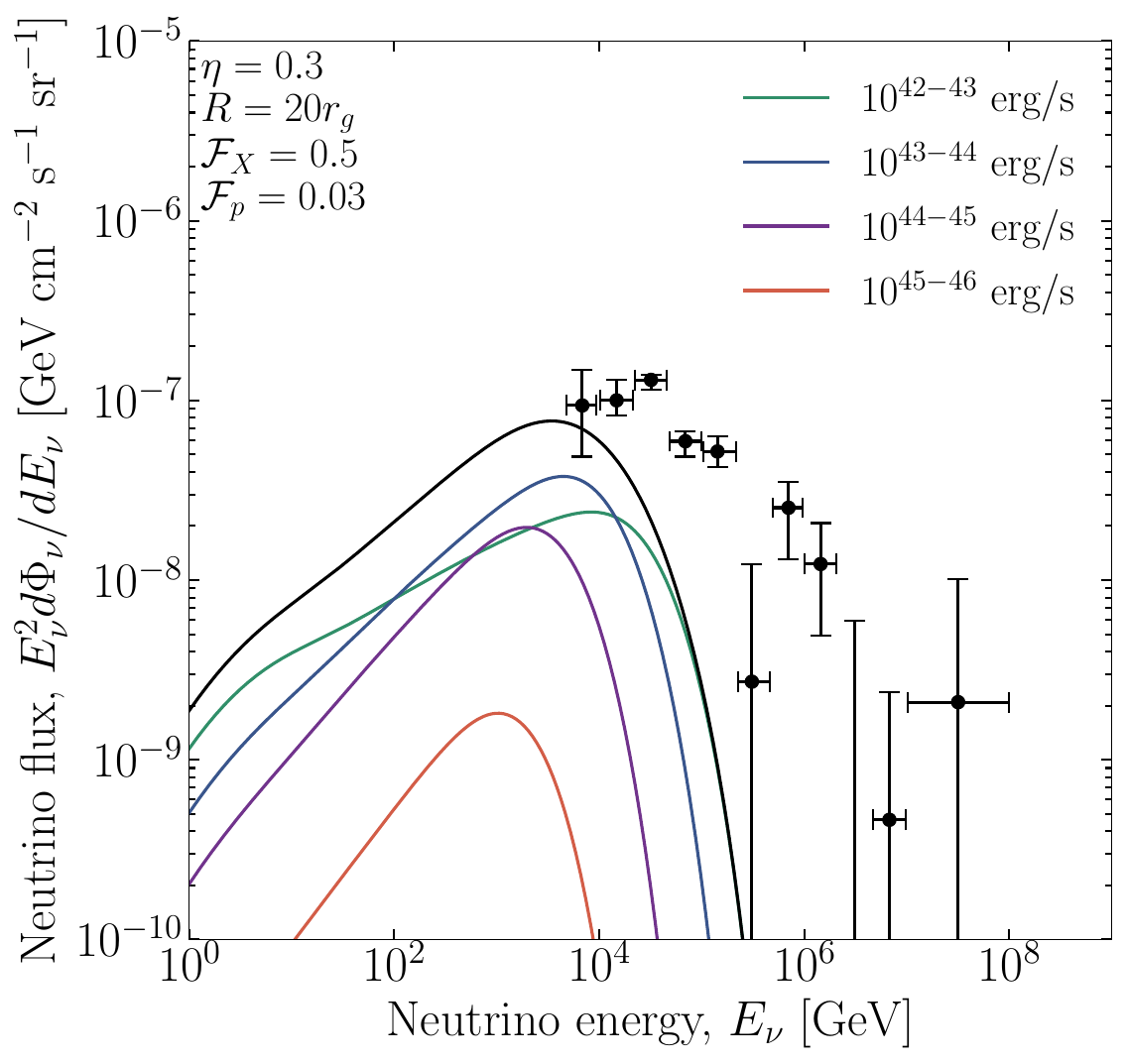}
    \caption{Same as Fig.~\ref{fig:diffuse}, but including the hydrodynamical escape.}
    \label{fig:diffuseh}
\end{figure}

It is also of interest to consider how the predictions for the diffuse neutrino flux are affected by the hydrodynamical escape. Fig.~\ref{fig:diffuseh} shows the corresponding results. Overall, the diffuse neutrino flux is lowered, but not by a significant amount. Generally, higher luminosity galaxies are less affected by this additional escape channel, since they are associated with larger values of $\sigma_{\rm tur}$ and therefore to a lower impact of the escape in comparison with the more rapid acceleration. The mild impact of the new escape channel is justified by the general dominance in the overall diffuse flux of Seyfert galaxies with luminosity comparable to or higher than that of NGC~1068, which as we have seen above is not strongly affected by the hydrodynamical escape.

\bibliographystyle{aasjournal}
\bibliography{References}

\end{document}